# Learning unknown pure quantum states

Sang Min Lee,[1,][*] Jinhyoung Lee,[2,][†] and Jeongho Bang[3,][‡]

[1]*Korea Research Institute of Standards and Science, Daejeon 34113, Korea*
[2]*Department of Physics, Hanyang University, Seoul 04763, Korea*
[3]*School of Computational Sciences, Korea Institute for Advanced Study, Seoul 02455, Korea*

We propose a learning method for estimating unknown pure quantum states. The basic idea of our method is to *learn* a unitary operation $\hat{U}$ that transforms a given unknown state $|\psi_\tau\rangle$ to a known fiducial state $|f\rangle$. Then, after completion of the learning process, we can estimate and reproduce $|\psi_\tau\rangle$ based on the learned $\hat{U}$ and $|f\rangle$. To realize this idea, we cast a random-based learning algorithm, called "single-shot measurement learning," in which the learning rule is based on an intuitive and reasonable criterion: *the greater the number of success (or failure), the less (or more) changes are imposed*. Remarkably, the learning process occurs by means of a single-shot measurement outcome. We demonstrate that our method works effectively, i.e., the learning is completed with a *finite* number, say $N$, of unknown-state copies. Most surprisingly, our method allows the maximum statistical accuracy to be achieved for large $N$, namely $\simeq O(N^{-1})$ scales of average infidelity. This result is comparable to those yielded from the standard quantum tomographic method in the case where additional information is available. It highlights a non-trivial message, that is, a random-based adaptive strategy can potentially be as accurate as other standard statistical approaches.

*Introduction.* — The characterization of a pure quantum state repeatedly generated from a preparation setup is a key step for many quantum applications [1, 2]. So far, these tasks have been performed with so-called "quantum state tomography (QST)" [3–5]. The conventional QST, which follows the standard statistical methodology, allows us to estimate unknown quantum states over a finite number $N$ of registered data from a set of measurement setups optimally chosen in advance. Such a standard approach is very appealing and has been the cornerstone of these practical tasks for decades, since it appears to be likely beneficial to the extraction of information from optimized measurements. However, it was proven in Ref. [6] that one can achieve $O(N^{-3/4})$ of average infidelity in standard (local) QST, whereas $O(N^{-1})$ is expected at best based on statistical grounds [33]. Thus, achieving higher accuracy, e.g., close to $O(N^{-1})$ infidelity, is still challenging both theoretically and practically.

Recently, it has been determined that achieving a level of accuracy at least as high as in the standard QST is possible using a different strategy, namely that of changing the measurements in an *adaptive* way. In this case, the measurement setting is appropriately chosen from trial to trial depending on the previously obtained measurement outcomes [10–14, 25]. Such adaptive QSTs have a number of practical advantages which include (i) the statistical errors are not as dominant; (ii) there is no need to deal with exponentially large data; (iii) (post) data analysis is not required [34]. Usually, the achievement of these advantages is established by the "optimal instructions" for the adaptive process. For example, one of the useful ways might be to use Bayesian estimation to decide the next-stage measurements [12, 17]. Quite recently, a variant of such adaptive strategies, called self-guided QST has been proposed with improved accuracy and efficiency [7, 18].

In this Letter, we propose an attractively simple and powerful method to estimate unknown pure quantum states. The main idea of our method is to *learn* a unitary operation $\hat{U}$ that transforms a given unknown state $|\psi_\tau\rangle$ to a known fiducial state $|f\rangle$. Then, after the learning is completed, we can infer and reproduce the unknown state $|\psi_\tau\rangle$ such that $|\psi_\tau\rangle \simeq |\psi_{\tau,\text{est}}\rangle = \hat{U}^\dagger |f\rangle$. To do this, we employ a novel learning algorithm, called "single-shot measurement learning (SSML)" [19, 20]. A significant and novel feature of SSML is that the learning proceeds based on the single-shot measurement outcomes. Thus, the practical advantages described in (i)-(iii) can also be achieved by invoking the adaptivity. In particular, we do not need to consider a large number of measurement setups, each of which is defined from a different observable quantity. Furthermore, we can expect little requirement of (classical) computational resources: e.g., no evaluations of the (in)fidelities are required at each learning step. The most remarkable result is that the average infidelity $\bar{\varepsilon} = 1 - \int d\psi_\tau \, |\langle\psi_{\tau,\text{est}}|\psi_\tau\rangle|^2$ scales $\simeq O(N^{-1})$ in our method. We show that this result is comparable to the yields from the standard QST in the case where additional information is brought.

*Scheme & method.* — We briefly describe how our method proceeds by specifying the key elements. Firstly, let us consider a preparation device (**P**) which can repeatedly generate unknown pure state $|\psi_\tau\rangle$ [35]. We also set a part of operation device (**U**) for the implementation of an arbitrary unitary $\hat{U}(\mathbf{p})$, where $\mathbf{p}$ is the vector whose components are *controllable* learning parameters. We then choose a fiducial state $|f\rangle$ freely, and let the measurement device (**M**) correspond to a "yes-or-no" ques-

tion, namely of whether we get the desired target:

$$\hat{M}_f = |f\rangle \langle f| \text{ and } \hat{M}_{f^\perp} = \hat{\mathbb{1}} - \hat{M}_f. \quad (1)$$

For convenience, the Hilbert-space dimension of $|f\rangle$ is assumed to be equal to $d$. Then, by connecting these three elements, we can define a system of the learning building-block, i.e., **P-U-M** (the "student" say), for conventional quantum information processing. In such a setting, we employ another key element which is the feedback system (**F**). It is responsible for the training (the "teacher" say). **F** has an optimal learning algorithm and a relatively small size of the (classical) memory to record the learning parameters. Then, the goal of the learning is to find a learning parameter vector $\mathbf{p}_{\text{est}}$ close to an optimal one in $\{\mathbf{p}_{\text{opt}}\}$, and finally estimate as:

$$\underbrace{|\psi_\tau\rangle = \hat{U}(\mathbf{p}_{\text{opt}})^\dagger |f\rangle}_{\text{Actual unknown-state}} \simeq \underbrace{|\psi_{\tau,\text{est}}\rangle = \hat{U}(\mathbf{p}_{\text{est}})^\dagger |f\rangle}_{\text{Estimated state}}. \quad (2)$$

Here, we note that the presented method can be referred to as a *quantum-classical hybrid* learning concept; i.e., the student is quantum and the teacher is classical. Such a hybridization would be easier and more economical to realize. There is also the possibility of gaining a quantum advantage from the quantum student [21, 22].

*Single-shot measurement learning (SSML).* — The efficiency and accuracy of our method strongly depends on the learning algorithm [36]. Here we employ a learning algorithm, called "single-shot measurement learning (SSML)" [37]. The intriguing and novel feature of the SSML is that the learner (i.e., $\hat{U}$ here) updates its own parameters by means of the single-shot measurement outcomes [38]. Specifically, the SSML runs as follows: For every learning step $n$, **P** generates $|\psi_\tau\rangle$ and it is transformed to an output state through **U**. Then, **M** performs the projective measurement with $\{\hat{M}_f, \hat{M}_{f^\perp}\}$ where each outcome is identified as a "success" or a "failure." More specifically, if a measurement result is $|f\rangle$, this is a success and regarded as one successful trial of the target task. Otherwise, we have a fail outcome. Thus we can infer that if the learning proceeds as expected, **M** will produce the more success outcomes; i.e., the number of *consecutive* successes, denote $M_S^{(n)}$, can be regarded as an index of how close the control parameters in $\mathbf{p}^{(n)}$ at the current $n$-step are to an optimal value $\in \{\mathbf{p}_{\text{opt}}\}$. As such, the rule for updating $\mathbf{p}$ is made as below:

[**R.1**] When we get a success outcome, **F** follows

$$M_S^{(n)} \leftarrow M_S^{(n-1)} + 1, \text{ and } \mathbf{p}^{(n+1)} \leftarrow \mathbf{p}^{(n)}. \quad (3)$$

At the first step, i.e., for $n = 1$, we set $\mathbf{p}^{(1)} \leftarrow \mathbf{r}$ and $M_S^{(0)} \leftarrow 0$, where $\mathbf{r}$ is a random vector whose components consist of random numbers.

[**R.2**] Otherwise, if the outcome is fail, **F** proceeds as:

$$M_S^{(n)} \leftarrow 0, \text{ and } \mathbf{p}^{(n+1)} \leftarrow \mathbf{p}^{(n)} + \omega \mathbf{r}, \quad (4)$$

where $\omega = \alpha(M_S^{(n-1)} + 1)^{-\beta}$ is the weight for the random vector $\mathbf{r}$. Here, $\alpha$ and $\beta$ are the free parameters related to the algorithm's performance.

Note that adopting the random vector $\mathbf{r}$ in [**R.2**], instead of using a pre-programmed one, is a typical strategy of machine learning [23], and is of particular importance in our method. These learning rules of the SSML—i.e., *the greater the number of success (fail), the less (more) changes are imposed*—intuitively makes sense.

The learning is not completed until $M_S^{(n)}$ becomes sufficiently large while producing no fail; more specifically, the learning is completed when the condition $M_S^{(n)} = M_H$ is met. We call this the "halting condition." After the learning is completed by satisfying this halting condition, we can obtain $\hat{U}(\mathbf{p}_{\text{est}})$ with $\mathbf{p}_{\text{est}} \leftarrow \mathbf{p}^{(n)}$. Here, the total iteration $n$ is the consumption $N$ of state copies for the estimation in Eq. (2). The learned $\hat{U}(\mathbf{p}_{\text{est}})$ is then expected to transform $|\psi_\tau\rangle$ to $|f\rangle$ faithfully, i.e., satisfying the following condition (for $M_H \gg 1$):

$$\varepsilon = 1 - \left|\langle f| \hat{U}(\mathbf{p}_{\text{est}}) |\psi_\tau\rangle\right|^2 = 1 - |\langle \psi_{\tau,\text{est}}|\psi_\tau\rangle|^2 \ll 1. \quad (5)$$

Note here that there exists a trade-off relation between inaccuracy and the learning time, depending on the predetermined number $M_H$; the larger (smaller) $M_H$, the lower (higher) infidelity $\varepsilon$ we have and the more (less) iterations or equivalently unknown-state copies in our case, are required to complete the learning process. Thus, it is very important to choose appropriate $M_H$ to account for the desired learning accuracy and time.

*Qubit-state estimation.* — To analyze our method, we here consider the estimation of unknown single-qubit state. Considering the possible realization of our approach, we adopt a general unitary learner, parameterized as

$$\hat{U}(\mathbf{p}) = \exp\left(-i\mathbf{p}^T \mathbf{G}\right), \quad (6)$$

where $\mathbf{p} = (p_x, p_y, p_z)^T$ is the control parameter vector and $\mathbf{G} = (\hat{\sigma}_x, \hat{\sigma}_y, \hat{\sigma}_z)^T$ is an operator vector whose components are SU(2) generators, i.e., Pauli operators. Note that $p_j$ ($j = x, y, z$) corresponds to the real hands-on control parameters, e.g., wave-plate angles for a polarization qubit in a linear-optical setup.

Firstly, we investigate whether our SSML method works well, i.e., whether the learning is completed in finite learning steps. To do this, we need to introduce the learning probabilities $P(N)$ defined as the probability that the learning is completed before or at a number $N$ of learning iterations [21]. Remarkably, the learning probability $P(N)$ is here analyzed as $\simeq 1 - \exp(-N/N_c)$ with a *finite constant* $N_c$ [39]. This means that in most case, learning is expected to be completed within a certain (i.e., $N_c$) learning steps. To verify this prediction, we performed numerical simulations: $10^4$ trials for each different halting condition $M_H$ [40]. We hereby note that the simulation is carried out, considering the linear-optical realization (see Sec. S1-A, S1-B, and S1-C of the Supplementary Information). The unknown states $|\psi_\tau\rangle$ are also randomly chosen for each trial. We extract the learning probabilities $P(N)$ from the obtained data and show that they are well fitted to the aforementioned function $1 - \exp(-N/N_c)$. Here, $N_c$ is estimated as $\simeq O(M_H^{0.869})$ (see Fig. 1). The results from the numerical analysis are in excellent agreement with our theoretical predictions.

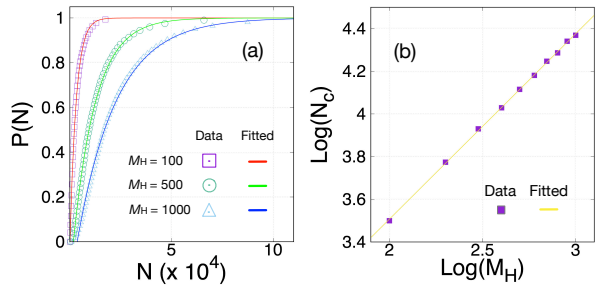

FIG. 1: (a) The learning probabilities $P(N)$ are drawn for $M_H = 100$, 500, and 1000. For each $M_H$, the data are obtained from $10^4$ estimation trials. In each trial, $|\psi_\tau\rangle$ is made at random. The data are well fitted to $1 - \exp(-N/N_c)$ with the factor $N_c$ which means the required number of state copies for the completion of the learning estimation. We get $N_c \simeq 3158$ for $M_H = 100$, $\simeq 13037$ for $M_H = 500$, and $\simeq 23377$ for $M_H = 1000$. (b) The graph of $N_c$ versus $N$ is also given on a log-log scale. The simulations are performed by increasing $M_H$ from 100 to 1000 at intervals of 100. Each data point of $N_c$ is obtained from $10^4$ simulations. By fitting the data, we get $N_c \simeq O(M_H^{0.869})$ (The detailed data are listed in Sec. S1-F of the Supplementary Information).

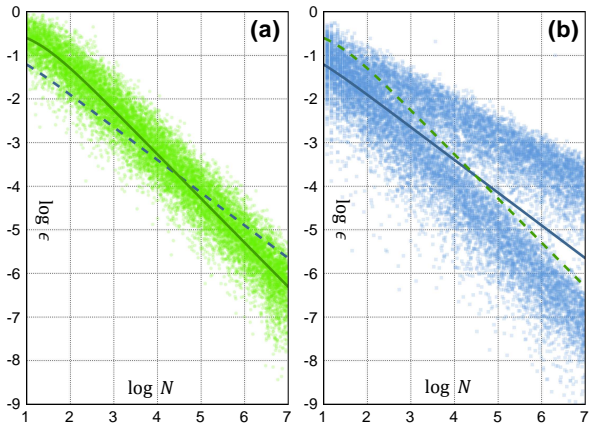

FIG. 2: The infidelities $\varepsilon$ are evaluated from (a) SSML and (b) SQST. We depict the graph of $\varepsilon$ versus $N$ as dots and their fitting lines on a log-log scale. The SSML result, i.e., $\bar{\varepsilon}_{\text{SSML}}$, exhibits the ultimate statistical accuracy, $O(N^{-\gamma})$ with $\gamma \simeq 1$ (green line), whereas $\gamma \simeq 0.75$ (blue line) for $\bar{\varepsilon}_{\text{SQST}}$. We note that we draw the (dashed and same colored) fitting lines in opponents for clearer comparison.

Secondly, we investigate the accuracy: i.e., the average infidelity $\bar{\varepsilon} = 1 - \int d\psi_\tau |\langle \psi_{\tau,\text{est}} | \psi_\tau \rangle|^2$ for large $N$. Noting that the adaptive estimators can be precise in a metrological scenario [22, 25–28], we expect that our SSML exhibits improved accuracy. To corroborate this, simulations are performed. The data from the standard QST (SQST) are also analyzed for comparison, where the observables are chosen from $\{\hat{\sigma}_x, \hat{\sigma}_y, \hat{\sigma}_z\}$ on each qubit. Figure 2 represents the results of our simulation in the form of $\varepsilon$ versus $N$ graphs on the log-log scale. By fitting the obtained data to $\varepsilon = C(N + N_0)^{-\gamma}$ [41], we evaluate the average infidelity $\bar{\varepsilon}$ with the main factor $\gamma$, such that $\bar{\varepsilon} \simeq O(N^{-\gamma})$. Here, it is determined that $\bar{\varepsilon}_{\text{SQST}} \simeq O(N^{-3/4})$ [42] and $\bar{\varepsilon}_{\text{SSML}} \simeq O(N^{-1})$. Most surprisingly, the result indicates that our SSML method is potentially better than the SQST approach. However, the SQST also exhibits improved accuracy, i.e., $\bar{\varepsilon}_{\text{SQST}} \simeq O(N^{-1})$, when the additional information— the fact that the unknown states are pure is used in the maximum-likelihood correction. This results thus support the idea that a learning estimation based on a random strategy is as efficient and accurate as the SQST (see also Refs [7, 11, 18, 24]).

*Summary.* — We have presented a simple but powerful method to estimate unknown pure quantum states $|\psi_\tau\rangle$. The main idea was to *learn* a unitary $\hat{U}$ to perform $|\psi_\tau\rangle \to |f\rangle$ for a known fiducial state $|f\rangle$. Then we could estimate $|\psi_\tau\rangle$, such that $|\psi_\tau\rangle \simeq |\psi_{\tau,\text{est}}\rangle = \hat{U}^\dagger |f\rangle$. To realize this idea, we casted a novel learning algorithm, called single-shot measurement learning (SSML), in which the learner ($\hat{U}$ here) was renewed according to a reasonable learning rule, i.e., the greater the number of success (fail), the less (more) adjustment is imposed. We noted that basically our method can be understood as a (weighted) random learning process with one-by-one measurements. As a main result, we demonstrated that our method works well for a finite number of state copies. Most surprisingly, we obtained higher accuracy, i.e., nearly $O(N^{-1})$ level of average infidelity, compared to $\simeq O(N^{-3/4})$ for the standard QST. However, we found that in the case where the additional information is available, the standard QST is also able to show $\simeq O(N^{-1})$ of average infidelity. This result implies an important and non-trivial scientific message, i.e., a random estimator can potentially exhibit high accuracy in quantum estimation which could be better than the maximum-likelihood estimator approach.

Our method brings is also associated with some operational advantages. Firstly, as the approach is an akin to the other adaptive approaches in that the advantageous features from the "adaptivity" can be carried over [29]. For example, it does not require excessive computational and experimental resources. Secondly, there is another operational advantage in that after the completion of learning, we can *directly* reproduce the estimated unknown state $|\psi_{\tau,\text{est}}\rangle$ even with no identification of the learned parameters in $\mathbf{p}_{\text{est}}$ [43]. This advantage is of particular significance, e.g., in a quantum cryptographic scenarios (see Refs. [30, 31]). We believe that our SSML method will find immediate application in quantum information tasks requiring pure state estimation.

*Acknowledgments.* – We are grateful to Jaewan Kim and Marcin Wieśniak for helpful discussions. SML and JB acknowledge the support of the R&D Convergence program of NST(National Research Council of Science and Technology) of Republic of Korea (No. CAP-18-08-KRISS). SML was also supported by KRISS projects (No. KRISS-2018-GP2018-0012, -0017). JL acknowledges the financial support of the Basic Science Research Program through the National Research Foundation of Korea (NRF) grant (No. 2014R1A2A1A10050117)

---

[43] In fact, we often meet such a situation. For example, consider a quantum linear-optical system, where the state of single-photon polarization is prepared and/or operated by a set of wave-plates [32]. Here, we note that it is impractical to identify the phase retardation of the wave-plates, precisely. Thus, one may have little confidence in the precise quantitative prediction of $\hat{U}$ from the identified $\mathbf{p}_{\text{est}}$, even though the learning is faithfully completed. Nevertheless, it is still guaranteed that the reproduced unknown-state $|\psi_{\tau,\text{est}}\rangle$ from the learned $\hat{U}$ is close to the actual unknown-state $|\psi_\tau\rangle$.

# Supplementary Information for "Learning unknown pure quantum states"

Sang Min Lee,[1,*] Jinhyoung Lee,[2,†] and Jeongho Bang[3,‡]

[1]*Korea Research Institute of Standards and Science, Daejeon 34113, Korea*
[2]*Department of Physics, Hanyang University, Seoul 04763, Korea*
[3]*School of Computational Sciences, Korea Institute for Advanced Study, Seoul 02455, Korea*

Supplementary Information for "Learning unknown pure quantum states."

## S1. DETAILS ON THE THEORETICAL AND NUMERICAL ANALYSIS OF SSML

### A. Polarization-based linear-optical experimental setting

To demonstrate that our method works well *in a real experiment* and with a high accuracy, we performed numerical simulations, particularly considering a polarization-based linear-optical realization (see Fig. S1). Firstly, we consider a single-photon source (SPS) and a combination of a quarter/half wave-plate (QWP/HWP) to construct **P**, by assuming that these elements are completely capsulated, e.g., in a black-box. Thus the single-photon state, $a\left|H\right\rangle + b\left|V\right\rangle$, generated in **P** is assumed to be an unknown state $\left|\psi_\tau\right\rangle$ (Here, $\left|H\right\rangle$ and $\left|V\right\rangle$ denote state of the horizontally and vertically polarized single-photon, respectively). Then, we employed a finite number of *controllable* wave-plates to implement **U**; i.e., the combination of QWP($\vartheta_1$)-HWP($\vartheta_2$)-QWP($\vartheta_3$), where the rotation angles $\vartheta_i$ ($i = 1, 2, 3$) consist of the control parameter vector $\mathbf{p} = (\vartheta_1, \vartheta_2, \vartheta_3)^T$. Here, note that these parameters $\vartheta_i$ ($i = 1, 2, 3$) are replaced by the general parameter vector, defined in Eq. (6) of the main manuscript. The wave-plate combination is the minimal requirement for an arbitrary single-qubit unitary operation [1]. The measurement **M** is implemented with the polarization beam-splitter (PBS) and two single-photon-detectors (SPDs). Here, for the sake of simplicity, we set $|H\rangle$ to be the fiducial state.

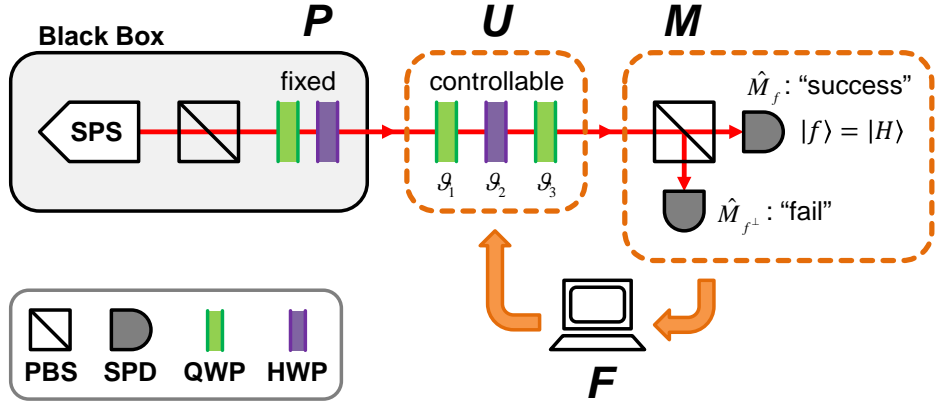

FIG. S1: A schematic layout for a linear-optical implementation of our SSML estimation.

[*]Electronic address: `samini@kriss.re.kr`
[†]Electronic address: `hyoung@hanyang.ac.kr`
[‡]Electronic address: `jbang@kias.re.kr`

### B. Number of wave-plate for U

The implementation of **U** can be performed by various combinations of wave-plates. Even though the minimum requirements of the wave-plate for an arbitrary unitary operation is QWP-HWP-QWP (QHQ), the combination of QH also can convert the fiducial state $|H\rangle$ to an arbitrary state. We thus investigate three types of wave-plate combinations: (a) QH, (b) QHQ and (c) QHQH. In the investigation, it is found that, theoretically, we can achieve $\gamma \simeq 1$ for all three cases with $\alpha \simeq 0.3$ and $\beta \simeq 0.5$ (see Fig. S2). Note however that the amount of phase retardation of the wave-plate is not ideal in practice. For this reason, QH would not allow a general transform. Note further that, as the number of wave-plates are increases, such imperfections will accumulate. Thus we utilized a QHQ setting in this work.

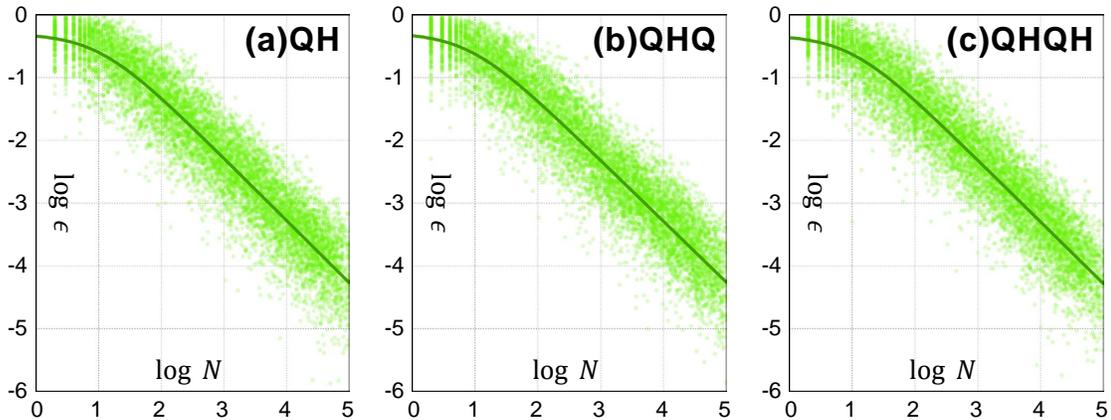

FIG. S2: The graphs of $\log N$ versus $\log \bar{\varepsilon}$ for (a) QH, (b) QHQ and (c) QHQH wave-plate combinations.

### C. State reproducibility of SSML: systematic error-insensitive

Some systematic errors, such as non-ideal phase retardation of the wave-plates, are inevitable in the real experiments. In particular, when we represent $|\psi_{\tau,\text{est}}\rangle$ from the final experimental setting $p_{\text{est}} = (\vartheta_1, \vartheta_2, \vartheta_3)_{\text{est}}^T$, the errors have an influence on the final estimation. However, when we reproduce the estimated state, our method is not affected by this error because it was confirmed by the experimental setup ($\mathbf{U}$) itself (satisfying the halting condition), and the state can be regenerated using the setup. More specifically, if we replace the measurement part (i.e., SPD for "success" side) to SPS, it naturally becomes the setting for the preparation of $|\psi_{\tau,\text{est}}\rangle$. Thus, we do not need to reconstruct the preparation setting with the identified (may be "poor") parameters. Such an advantage can not be found in SQST and other existing AQST schemes, since those methods implicitly assume that the theoretical description and the implementation of the experiment are perfectly matched.

### D. Optimization of the SSML parameters $\alpha$ and $\beta$

In order to optimize the feedback range $\omega$ in Eq. (4) of the main text, we examine the relationship between $\varepsilon$ and $M_S$. To do this, let us assume that $\mathbf{p}$ is near to an optimal $\mathbf{p}_{\text{opt}}$, but still not sufficient to complete the algorithm. Noting that $F = \left|\langle f|\hat{U}(\mathbf{p})|\psi_\tau\rangle\right|^2$, the probability that we get the number $M_S$ of successes continuously is given as $p(M_S) = F^{M_S}(1-F)$. Thus, we have

$$\overline{M_S} = \frac{F}{1-F} \simeq \frac{1}{1-F} = \frac{1}{\varepsilon}. \tag{S1}$$

Because $F$ is close to 1 with $\mathbf{p} \simeq \mathbf{p}_{\text{opt}}$, the infidelity $\varepsilon$ is approximated as $\simeq A(\mathbf{p} - \mathbf{p}_{\text{opt}})^2$. Then, the distance between $\mathbf{p}$ and $\mathbf{p}_{\text{opt}}$ is represented as $|\mathbf{p} - \mathbf{p}_{\text{opt}}| \propto \overline{M_S}^{-1/2}$. The parameter $\beta$ is thus approximately 0.5, which is in good agreement with the simulation results. However, $\alpha$ should be found in a heuristic manner. In this case, we found that the optimal setting is $\alpha \simeq 0.3$.

### E. Learning probability and the effectiveness of SSML

Here we approximately estimate $P(N)$ by using the random learning strategy, which is often casted for the analysis. To this end, we first consider the probability $p_s = (1-\varepsilon)^{M_H}$ that the learning is completed for $\mathbf{p}_{\text{est}} \simeq \mathbf{p}_{\text{opt}}$. Here, $1-\varepsilon$ is the probability of the success event, namely that of measuring the fiducial state $|f\rangle$. Then, we introduce a continuous function,

$$\frac{1}{2} \leq \Xi(\mathbf{p}) = \xi_1(p_1)\xi_2(p_2)\xi_2(p_3) \leq 1, \tag{S2}$$

satisfying $\Xi(\mathbf{p} \neq \mathbf{p}_{\text{opt}}) < \Xi(\mathbf{p}_{\text{opt}}) = 1$. We note that this function $\Xi(\mathbf{p})$ is obtained by minimizing $\left| p_s^{1/M_H} - \Xi(\mathbf{p}_{\text{opt}}) \right|$. Thus we can assume that $p_s = (1-\varepsilon)^{M_H} \simeq \Xi(\mathbf{p}_{\text{est}})^{M_H} \simeq \Xi(\mathbf{p}_{\text{opt}})^{M_H}$ *for very large* $M_H$. We then use an interesting idea by approximating $\xi_j(p_{j,\text{est}})^{M_H}$ ($j = 1, 2, 3$) with a delta function,

$$\xi_j(p_{j,\text{est}})^{M_H} \approx \exp\left[-\frac{(p_{j,\text{est}} - p_{j,\text{opt}})^2}{\sigma^2}\right]. \tag{S3}$$

In the circumstance, we estimate the average probability $\overline{p_s}$, such that (for $\sigma \ll 1$ [2])

$$\overline{p_s} \simeq \int dp\, \xi_1(p_{1,\text{est}})^{M_H} \int dp\, \xi_2(p_{2,\text{est}})^{M_H} \int dp\, \xi_3(p_{3,\text{est}})^{M_H} \approx (\sqrt{\pi}\sigma)^3. \tag{S4}$$

where $\sigma$ is the value of the deviation of $p_{j,\text{est}}$ about the optimal $p_{j,\text{opt}}$. Here, the approximation of the right part of Eq (S4) is made with the assumption that the space of $(p_{1,\text{est}}, p_{2,\text{est}}, p_{3,\text{est}})$ is isotropic [3]. Then, for any sequence $\mathbf{p}^{(0)} \to \mathbf{p}^{(1)} \to \mathbf{p}^{(2)} \to \cdots \to \mathbf{p}^{(N)} = \mathbf{p}_{\text{est}} \simeq \mathbf{p}_{\text{opt}}$ of updating the parameter vectors in the learning process, we can approximate the learning probability $P(N)$ as

$$\begin{aligned}
P(N) &\approx \overline{\Xi(\mathbf{p}^{(1)})^{M_H}} \\
&+ \left(1 - \overline{\Xi(\mathbf{p}^{(1)})^{M_H}}\right) \overline{\Xi(\mathbf{p}^{(2)})^{M_H}} \\
&+ \left(1 - \overline{\Xi(\mathbf{p}^{(1)})^{M_H}}\right) \left(1 - \overline{\Xi(\mathbf{p}^{(2)})^{M_H}}\right) \overline{\Xi(\mathbf{p}^{(3)})^{M_H}} \\
&\vdots \\
&+ \left(1 - \overline{\Xi(\mathbf{p}^{(1)})^{M_H}}\right) \left(1 - \overline{\Xi(\mathbf{p}^{(2)})^{M_H}}\right) \cdots \left(1 - \overline{\Xi(\mathbf{p}^{(N-1)})^{M_H}}\right) \overline{\Xi(\mathbf{p}^{(N)})^{M_H}} \\
&\approx \sum_{k=0}^{N-1} \left(1 - \overline{\Xi(\mathbf{p}^{(k)})^{M_H}}\right)^k \overline{\Xi(\mathbf{p}^{(N)})^{M_H}}.
\end{aligned} \tag{S5}$$

Here, assuming that the learning process is started with a parameter vector $\mathbf{p}^{(0)}$ close to $\mathbf{p}_{\text{opt}}$ [4], we can assume $\overline{p_s} \simeq \overline{\Xi(\mathbf{p}^{(j)})^{M_H}}$ for all $j = 0, 1, \ldots, N$. Then, we finally arrive at

$$P(N) \approx 1 - (1 - \overline{p_s})^N = 1 - e^{-\frac{N}{N_c}}, \tag{S6}$$

for very large $N$. Here, $N_c \simeq \overline{p_s}^{-1} \simeq (\sqrt{\pi}\sigma)^{-3}$, which is the average number of iteration to complete the learning. This also indicates that we need a large iteration to achieve more accurate learning.

### F. Detailed simulation results of the learning probability

We describe detailed simulation results of the learning probability in this subsection. In particular, in order to investigate whether or not the learning is completed in a finite number of learning steps as predicted in the previous subsection, we analyze the learning probability $P(N)$, which is defined as the probability that the learning is completed before or at a number $N$ of learning iterations. Here, let us recall the fact that because $P(N)$ is an cumulative distribution, the constant factor $N_c$ in Eq. (S6) can be interpreted as the average number of iterations for the completion of the learning. Having the aforementioned in mind, we perform numerical simulations for analysis. The simulations are performed for different halting conditions (from $M_H = 100$ to $M_H = 1000$ at intervals of 100, total 10 cases). For each case of $M_H$, we perform $10^4$ simulations to construct the learning probability $P(N)$. In Fig. S3, we draw $P(N)$ for each case of $M_H$. The data are well fitted to the function in Eq. (S6). The values of $N_c$ found from the data fitting are listed in Tab. S1. The $N_c$ values are well matched to the average iterations $\overline{N}_\text{data}$ evaluated from the actual simulation data. Our analytical predictions in the previous subsection are thus well borne out.

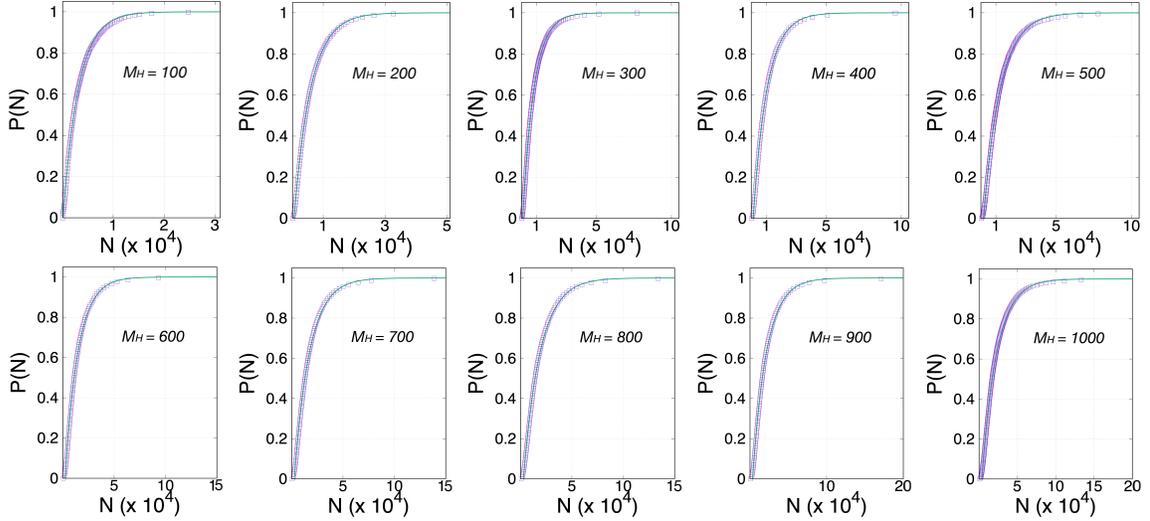

FIG. S3: The learning probabilities obtained from the SSML simulations. The simulations are performed for different halting conditions (from $M_H = 100$ to $M_H = 1000$ at intervals of 100, totally 10 cases).

| Halting condition $M_H$ | $N_c$ ($\overline{N}_\text{data}$) | Halting condition $M_H$ | $N_c$ ($\overline{N}_\text{data}$) |
|---|---|---|---|
| 100 | $\simeq 3158$ ($\simeq 3354$) | 600 | $\simeq 15162$ ($\simeq 15240$) |
| 200 | $\simeq 5942$ ($\simeq 6096$) | 700 | $\simeq 17657$ ($\simeq 17777$) |
| 300 | $\simeq 8513$ ($\simeq 8780$) | 800 | $\simeq 19327$ ($\simeq 19464$) |
| 400 | $\simeq 10692$ ($\simeq 10951$) | 900 | $\simeq 21915$ ($\simeq 22112$) |
| 500 | $\simeq 13037$ ($\simeq 13255$) | 1000 | $\simeq 23377$ ($\simeq 23381$) |

TABLE S1: The detailed values of the fitting parameter $N_c$ are listed for each $M_H$. The value $\overline{N}_\text{data}$ are obtained by averaging over $10^4$ data of simulations.

## S2. FURTHER INVESTIGATION WITH AN ENSEMBLE-BASED LEARNING

Using Eq. (S1), we can determine the updating range $\omega = \alpha(M_S + 1)^{-\beta}$. However, since $\Delta M_S = \sqrt{F}/(1 - F)$ obtained from $p(M_S)$ is very large when $F \simeq 1$, the determined value of $\omega$ can be considered to be unreliable. Thus, one can consider the learning via an ensemble measurement that deals with a number $M_E$ of samples, instead of a single-shot under the same setup $\mathbf{p}^{(n)}$. In particular, such an ensemble-based learning can be considered to be more accurate and efficient than the single-shot, as the fluctuation $\Delta M_S$ is very small when $F \simeq 1$. For the ensemble measurements, $M_S$ represents the total number of success in the single parameter $\mathbf{p}^{(n)}$, rather than consecutive successes. To test the ensemble-based measurement learning (EML), we perform numerical simulations according to the following rules: firstly, $M_E$ copies are measured at $n$-th learning step. Then, $\mathbf{F}$ updates $\mathbf{p}^{(n)}$ by using the number $M_S^{(n)}$ of measurement results, such that

$$\mathbf{p}^{(n+1)} \leftarrow \mathbf{p}^{(n)} + \alpha \left( \frac{M_E - M_S^{(n)}}{M_E} \right)^\beta \mathbf{r}, \tag{S7}$$

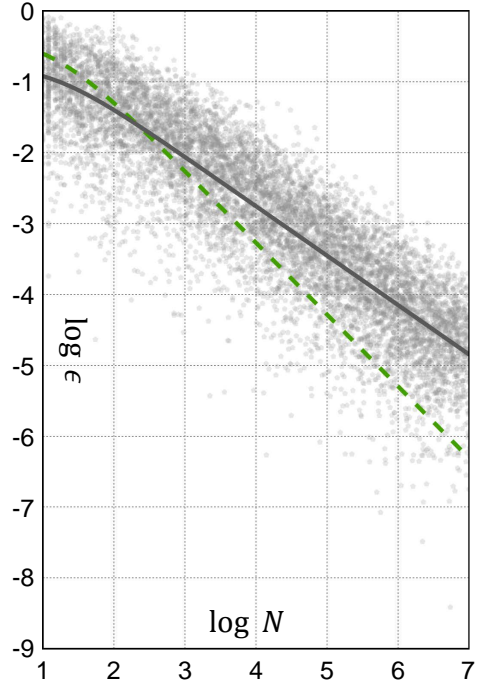

FIG. S4: The graph of $N$ versus $\varepsilon$ on the log-log scale for EML results. The data are fitted to $\bar{\varepsilon} \simeq O(N^{-\gamma})$ with $\gamma \simeq 0.70$ (gray line). For comparison, we also draw the line (dashed green) of SSML. It is directly observed that the SSML is superior than EML.

until $M_S^{(n)}$ becomes equal to $M_E$. The EML simulations are performed by varying the parameters $\alpha$ and $\beta$. In Fig. S4, we present the best results. Here, we get $\bar{\varepsilon} \simeq O(N^{-0.7})$. Clearly, the result is inferior to those of the SSML method. The reason is because even when $F \simeq 1$, one can arrive at the situation $M_S^{(n)} > M_S^{(n+1)}$, because the parameter update, i.e., learning, is performed by the random vector $\mathbf{r}$. Thus, the resources of the state-copy do not need to be wasted when $F$ is not close to 1. Noting the aforementioned, one can infer that it is an optimal (i.e., resource efficient) strategy to carry out the (single-shot) measurements until the failure event appears.

### S3. MORE DETAILS ABOUT THE STANDARD QST SIMULATIONS

In the SQST simulations, we can assort the data into the two groups, as shown in Fig. S5: (a) The red point data are reconstructed to the mixed states (geometrically, inside the Bloch sphere) without the maximum-likelihood (ML) method. (b) On the other hand, the (raw) data included in the other group, denoted by the purple points, are initially not the legitimate physical states (i.e., outside the Bloch sphere). Therefore, we need to correct the data using the ML method, so that they are transformed to the pure state (i.e., on the Bloch sphere). Thus, the fidelities evaluated from (b) are higher than those from (a) in general. Fitting the data corresponding to each case (a) and (b), we can find that $\bar{\varepsilon} \simeq O(N^{-\frac{1}{2}})$ (red solid line) and $\bar{\varepsilon} \simeq O(N^{-1})$ (purple solid line). The ratio of the amounts of data corresponding to (a) and (b) is nearly 50:50. Therefore, in a realistic application, we will observe $\bar{\varepsilon} \simeq O(N^{-3/4})$ (blue dashed line) from every data. Nevertheless, if we use the additional information, i.e., the fact that the unknown state is pure in our case, the ML method moves all states to the pure state (i.e., the data in (a) are transferred into (b)), and the overall data are equal to the case of (b); namely, we get $\bar{\varepsilon} \simeq O(N^{-1})$ (purple solid line).

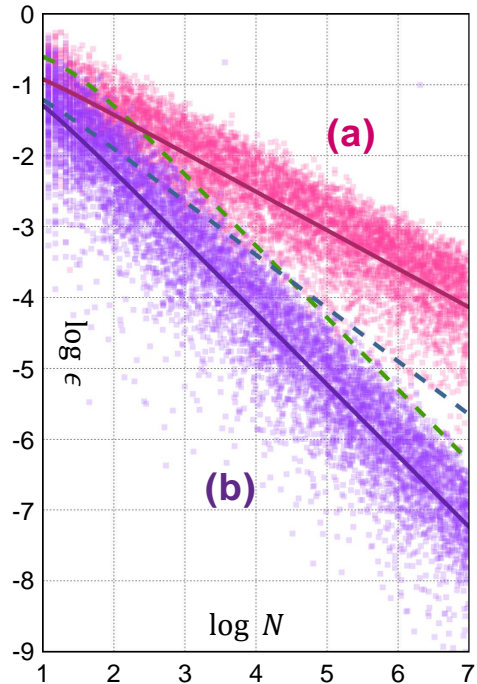

FIG. S5: The Fig. 2 in the main manuscript is redrawn, where the data are sorted into the two groups, each of which is characterized from those of the (a) mixed and (b) pure estimated states (see the text).

---